European Society for Astronomy in Culture
SEAC Publications; Vol. 01

Sonja Draxler, Max E. Lippitsch
& Gudrun Wolfschmidt

# Harmony and Symmetry

Celestial regularities shaping human culture

Proceedings of the SEAC 2018 Conference in Graz

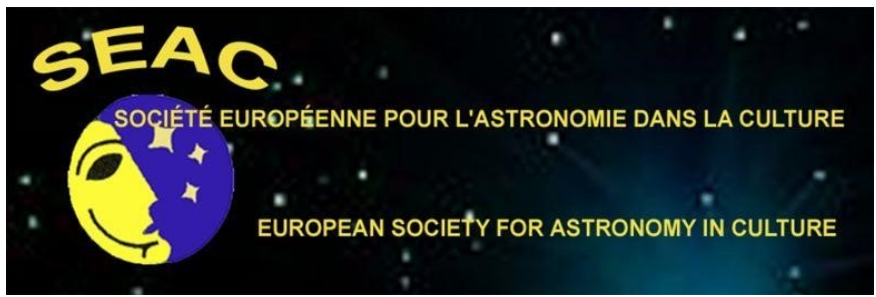

Hamburg: tredition 2020

# Calendrical Interpretation of Spirals in Irish Megalithic Art

*Marc Türler*

*Independent scholar, formerly at the Department of Astronomy of the
University of Geneva*

*marc.turler@unige.ch, marc.turler@gmail.com*

**Abstract:** The tumuli of Newgrange and Knowth in Ireland are among the most monumental heritages of the Neolithic era. The megalithic constructions date back to around 3'200 BC, centuries before the completion of Stonehenge and the Egyptian pyramids. Passageways inside the mounds have been shown to be aligned such as the rising sun illuminates the interior chambers on the winter solstice at Newgrange and around the equinoxes at Knowth. Many of the kerb and interior stones are covered with petroglyphs, in particular with spiral motifs. Despite several attempts to classify and interpret them, they remain enigmatic. Here we show that some of the most elaborated spirals are likely calendrical representations. We use a new, 'dynamic' approach adding a temporal dimension to the rock art. We identify in the detailed spiral motifs up to five different manifestations of the solar and lunar cycles, which could easily be observed in prehistoric Ireland. Corroborating evidence comes from adjacent motifs or a special location of the stone. Although the work is speculative by nature, many clues give confidence in the interpretation. If correct, it would imply that one or a few individuals developed skills for celestial observations, the design of complex motifs and their preservation on the rock. This resembles a scientific process, but is more likely the expression of a ritual engagement with the sky.

**Keywords:** archaeoastronomy; calendars; cycles; gnomon; Ireland; Knowth; Neolithic; Newgrange; petroglyphs; rock art; spirals; symbols

## Introduction

Newgrange and Knowth are two large megalithic mounds both located in the bend of the river Boyne not far from Dublin. Covered with grass and encircled by kerbstones, they used to be considered as graves and called passage tombs, because one or several corridors lead to interior chambers. Although there is evidence for burial human remains (O'Kelly, 1982; Eogan, 1986), the astronomical orientation of the passageways suggests that these monumental constructions were not primarily tombs (e.g. O'Callaghan, 2004; Murphy, 2012). Their design aims to let sun rays illuminate the chamber of Newgrange at sunrise on the winter solstice (Patrick, 1974; Ray, 1989) and the two corridors of Knowth at sunrise and sunset near the time of the equinoxes (Eogan, 1986; Ruggles, 1999; Prendergast & Ray, 2017).

### *On the abstract nature of the megalithic art*

Many of the kerb and interior stones at Newgrange and Knowth are covered with petroglyphs (O'Kelly, 1982; Eogan, 1986). This megalithic art is dominantly non-figurative and often geometric, although there may be rare exceptions of anthropomorphic figures and elements of figuration within the abstract art (O'Sullivan, 1997; Robin, 2012). Such an art could be purely ornamental without any meaning, but one would then expect more repetition in the motifs. On the contrary, every decorated stone has a unique composition of simple elements such as triangles, lozenges, arcs, concentric circles, spirals, chevrons, and wavy lines (e.g. Shee Twohig, 1981).

In the early 1990s emerged the idea that the megalithic art could be related to visions during altered states of consciousness (Lewis-Williams and Dowson, 1993; Dronfield 1995). This was deduced from the analogy of some of the prehistoric motifs with drawings performed when taking drugs. However, whereas drawing dots, zigzags and lines on a sheet of paper can be done in such a state of mind, engraving them on the rock with stone tools seems almost impossible. Furthermore, the complexity of some patterns make them unlikely to reflect a vision in a state of altered consciousness. The effort put into the conception and painful engraving of complex symbols, such as double and triple spirals, suggests a higher intention than either decoration or recording of hallucinations.

### *The art most likely represents celestial manifestations*

In the quest of the hidden meaning of the abstract art, the most promising route seems to be an association with celestial manifestations. Indeed, the regular cycles of the sun, the moon and the stars must have fascinated mankind already in prehistoric times. At least in the Neolithic era, settled communities would most likely have attached a great importance to these cycles, because they are linked to the passage of seasons affecting their agricultural life. Another clue is that the most richly engraved stones are not placed randomly, but are oriented towards specific celestial points. This is particularly evident in Newgrange with kerbstones K52 and K67 oriented towards the direction of sunrise and sunset at the summer solstice (Brennan 1983: 192). Another hint is the presence of several fan-shape petroglyphs evoking sundials, in particular at Knowth on the flat top of kerbstone K7 and forming the central motif of kerbstone K15, sometimes called the 'sundial stone' (Brennan, 1983; Harley, 2009). Finally, the intentional alignment of the passageways with the direction of sunrise or sunset on specific days of the year is a clear evidence of the importance of annual cycles to this Neolithic community. In such a context, it would be natural that at least part of the art is also related to astronomical manifestations. This possibility was explored intensively by Brennan (1983) and Thomas (1988), who presented interesting ideas, some of them rather convincing, others much less. The most explicit representation of astronomical cycles seems to be on kerbstone K52 at Knowth. This so-called 'calendar stone' shows a succession of 22 crescents or horseshoes completed by 7 circles. Their arrangement in an ellipsoid leaves little doubt that they represent the moon phases over 29 days, i.e. the synodic month (Brennan, 1983; Prendergast, 2017). The associated wavy line has further been successfully interpreted by Brennan (1983) as a count of months to synchronise the lunar and annual cycles. Besides this example and possibly some other less robust claims, the vast majority of the rock art could not yet been convincingly and coherently interpreted.

This article aims to reconsider the idea that a fraction of the megalithic art is a representation of astronomical manifestations. This is done with a new, 'dynamic' approach of the art, which is shown in the next section to be promising. We then present a set of observable properties of astronomical cycles, which could have been easily observed by Neolithic people. For the actual interpretation, we focus our analysis on spiral motifs of the Bend of the Boyne, which seem to have been at the origin of a very enduring pictorial tradition (Morrison, 2005). There are several reasons for this choice: 1) spirals are very typical of the art at Newgrange and are also often present at Knowth; 2) a spiral traces a series of circles alike the apparent movement of the celestial bodies in the sky; and 3) the spiral motif seems to be the most complex design of the megalithic art with many variations from the simplest spirals to complex S-shaped, double and triple spirals. For a set of five specific spiral motifs, we identify characteristics of the design, which appear to match the seasonal, celestial manifestations of the sun and the moon. We then discuss the validity of the interpretation and put the study in context with respect to other spiral petroglyphs and earlier work.

**A dynamic approach of the art**

The difficulty until now to interpret the art might have been to look at it in a static way, simply as two dimensional pictures. The creation of the motifs is instead a dynamic process. Including this temporal component could reveal new insights. For instance, when you represent a filled triangle (▲) or a filled lozenge (♦), there is more material to remove with a series of pick-marks at the centre of these geometric figures than on their left and right sides. The engraving process therefore takes more time at the centre. A dynamic interpretation of both signs is that they could represent a variable period of time with an increase followed by a decrease in duration. Such signs could for instance represent the variable length of daylight from winter to summer and back to winter. More generally, they could express the abstract notion of increase followed by decrease or growth followed by decay, as can be observed in nature: sunrise and sunset, waxing and waning moon phases, rising and ebbing tides. Symbolically, this could also reflect the cycle of life with the birth and death of flowers, trees, animals, and humans. This special meaning would explain why triangles and lozenges are so much more frequent in Irish megalithic art than other geometric forms such as squares and rectangles (e.g. Shee Twohig, 1981).

Similarly, it is not enough to just look at a complex spiral motif, but one really needs to follow its pattern with the finger to feel the longer time it took the artist to engrave them as the circles enlarge away from the centre. In a dynamic interpretation this could be a representation of the abstract notion of an increasing duration over several cycles. A typical example is the seasonal increase of daylight over several lunar cycles. The frequent association of filled triangles or lozenges with spirals, which is very obvious on all three richly decorated kerbstones of Newgrange (K1, K52, and K67), would find an explanation with such a dynamic approach of the art. All three motifs could represent the abstract notion of a duration changing with time, such as daytime over a year.

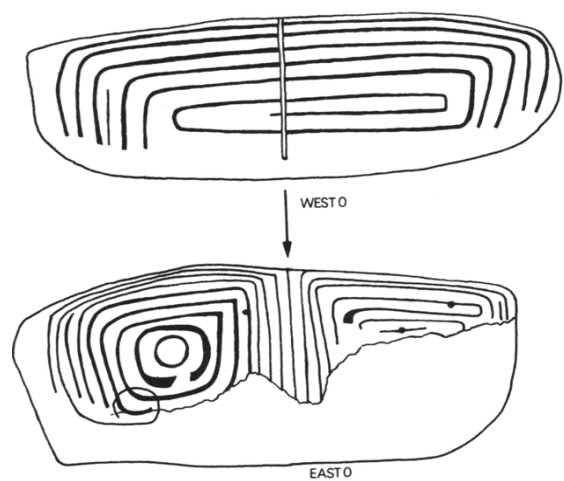

**Figure 1.** The two kerbstones K11 (bottom) and K74 (top) respectively in front of the east and west entrances of the main mound of Knowth (from Brennan, 1983). They are plausibly representing two complementary characteristics of the equinoxes.

*An interpretation of the entrance stones of Knowth*

The pattern engraved on the kerbstone K74 in front of the west entrance of Knowth looks like a simple decoration (Figure 1). Can we find out something more by following the groove? The motif has a central vertical line indicating the entrance of the passage and also the equinox, because the shadow of a narrow standing stone is cast onto it at sunset around the equinox (Brennan, 1983: 104). A horizontal line starts from there towards the right, stops and continues upwards for a while before coming back to the left crossing the equinoctial, vertical line and going the same length in the other direction before stopping again moving downwards and returning to the right. The rest of the motif, although quite altered nowadays, seems to be a continuation of this pattern with horizontal passages through the vertical line and stops on both sides. There is a very striking parallel between this movement and the varying position of the sunset over the year. Starting from the west at the autumn equinox, the position of sunset rapidly moves towards the south then stops at the winter solstice and moves back to cross the west direction at the spring equinox and continues to move north for three months before stopping at the summer solstice and reverting again direction towards the south, and so on. The theme of this decoration is plausibly the annual, pendular movement of the sunset from south-west to north-west.

The opposite kerbstone K11 at the entrance of the east passage has also a central vertical line, but here there are two separate sets of concentric shapes located symmetrically on both sides (Figure 1). The motif is roughly circular at the centre and becomes more boxy towards the border, probably to better match the rectangular shape of the stone. The full design seems to be a stylised representation of the gnomon pattern (see Figure 4), thus possibly emphasising the second aspect of the equinox, which is the east-west division of the gnomon pattern (see below). The two entrance stones would thus represent the two complementary aspects of the equinox: a division of the year in a dark autumn-winter season and a light spring-summer season, and a point of regular passage from the one to the other.

**Observable properties of astronomical cycles**

*The main astronomical cycles*

Days, months and years are observable astronomical cycles. The day is due to the spin of the Earth around its north-south axis and results in an apparent movement of the sun, the moon, and the stars. The month relates to the revolution of the moon around the Earth. The sidereal month (~27.32 days) is the time the moon takes to complete a full revolution through the constellations, while the synodic month (~29.53 days) is the more familiar interval between two consecutive full moons. Both phenomena can be followed easily with the naked eye, it just requires regular observations. The (tropical) year (~365.24 days) relates to the orbit of the Earth around the sun in the ecliptic plan. As seen from Earth, this results in the annual rotation of the sun through the constellations of the zodiac. The tilt of ~23.5 degrees between the spin axis of the Earth and its rotation axis around the sun causes the seasons and associated celestial manifestations described below.

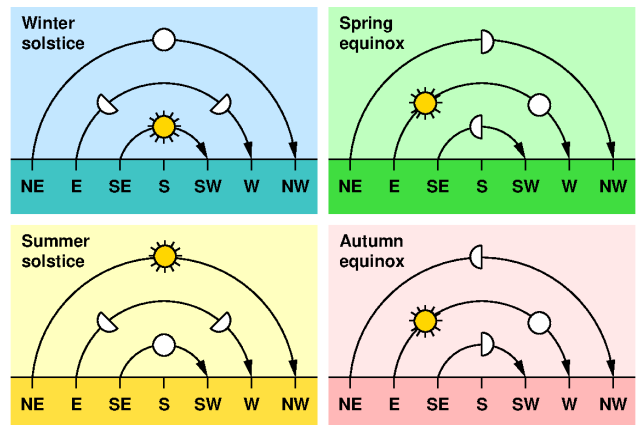

**Figure 2.** Schematic representation of the daily paths of the sun and the moon at solstices and equinoxes for a typical latitude of ~45° N and ignoring the small disturbance due to the ~5° inclination of the moon orbital plan with respect to the ecliptic. The full moon (white circle) and the sun (rayed yellow circle) follow opposite extreme paths on the winter and the summer solstices. The wide, 'dominant' path from north-east (NE) to north-west (NW) is followed by the first-quarter moon on the spring equinox and by the third-quarter moon on the autumn equinox.

*The calendrical challenge*

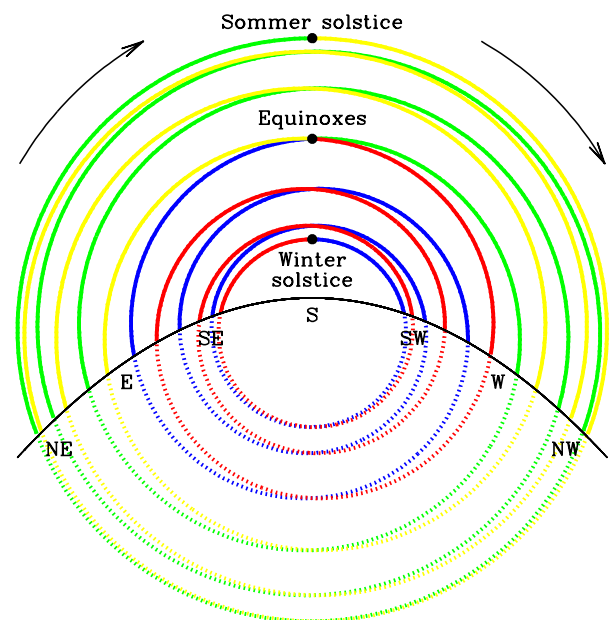

**Figure 3.** Schematic representation of the apparent spiralling path of the sun in the sky and its extrapolation below the horizon (dotted line) during a year. For the sake of clarity, we draw only one apparent turn of the sun per month instead of one per day. The horizon is shown in perspective with cardinal points placed for a location at a medium northern latitude (~45° N). Solstices and the equinoxes are shown with dots, seasons with colors. The line spirals outwards in winter (blue) and spring (green) before returning inwards in summer (yellow) and autumn (red).

In a simple world, the month would have an integer number of days and the year an integer number of months. As this is not the case – a year counts ~12.37 synodic months – finding a coherent relation between the three time units is a puzzle since millennia. A very simple calendar is the ancient Egyptian civil calendar dividing the year into 12 months of 30 days plus five additional days (Richards, 1998: 153). Working only with integers, a closer match is however reached with 13 months of 28 days totalising 364 days, which needs just one additional day to complete the year. There are several arguments in favour of such a calendar. Indeed, 28 days in a month is in between the sidereal and synodic cycles, corresponds to the number of double-cycles of tides (~24 hours and 50 minutes) in a synodic month, and matches on average the menstruation cycle. The number 28 is furthermore divisible by four to mark quarter moons, our modern weeks. Although there is no such calendar currently in use, several primitive calendars have 13 months, like those of some of the Native American tribes, e.g. Lakota and Cherokee, possibly the Rapa Nui calendar of the Easter Island, as well as the Celtic Tree Calendar. A 13-month calendar was also proposed in 1849 by August Comte and called the 'positivist calendar' (Richards, 1998: 113) and the idea was taken up in the early 20th century as the perennial International Fixed Calendar. On the other hand, the now commonly used 12-month calendar has the great advantage to be divisible in two equal semesters and in four seasons of three months each.

*Seasonal celestial manifestations*

Besides the calendrical challenge to relate daily, monthly, and yearly cycles, there is a series of interesting seasonal manifestations to be observed by people living at relatively high geographic latitude. The most affecting consequence is that the night is much longer in the winter than in the summer and this is directly related to the daily path of the sun in the sky. In winter, it traces only a small arc on the southern horizon, while in summer it draws a long, high arc from north-east to north-west (see Figure 2). Starting from the winter solstice, the solar path enlarges on the horizon over days and months until the summer solstice and then diminishes again in the second part of the year. By understanding that the sun has to return from west to east below the horizon during the night, the changing width of the daily movement of the sun from one solstice to the other naturally results in a spiral pattern (Brennan, 1983: 188). Figure 3 schematically illustrates this spiralling pattern as it unfolds day after day over a year.

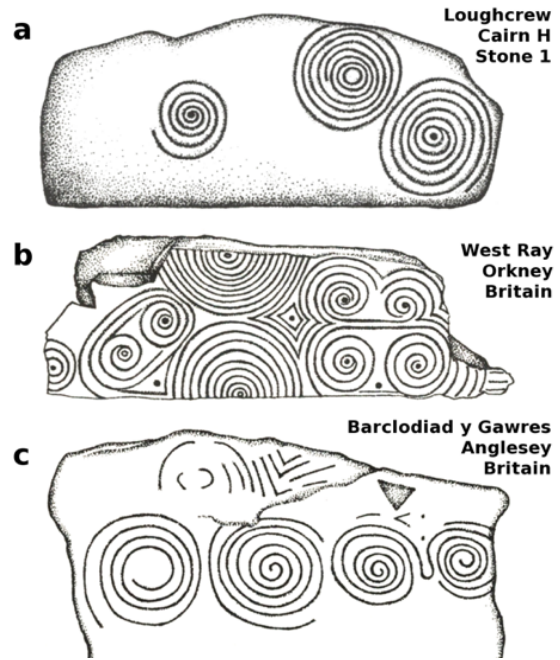

**Figure 5.** Three examples of stones engraved with one or several pairs of rings, cup-and-rings and/or spirals, which are likely inspired by the gnomon pattern. The drawings are all from Brennan (1983): **a.** chamber stone 1 in Cairn H at Loughcrew, Ireland with a pair of adjacent spirals with central rings and, interestingly, only one cup mark; **b.** fully engraved stone at West Ray, Orkney, Britain with a pair of concentric semicircles and three other pairs of connected spirals; **c.** another variation on the theme are the two pairs of spirals – one with a peculiar central connection – on a stone from Barclodiad y Gawres, Anglesey, Britain.

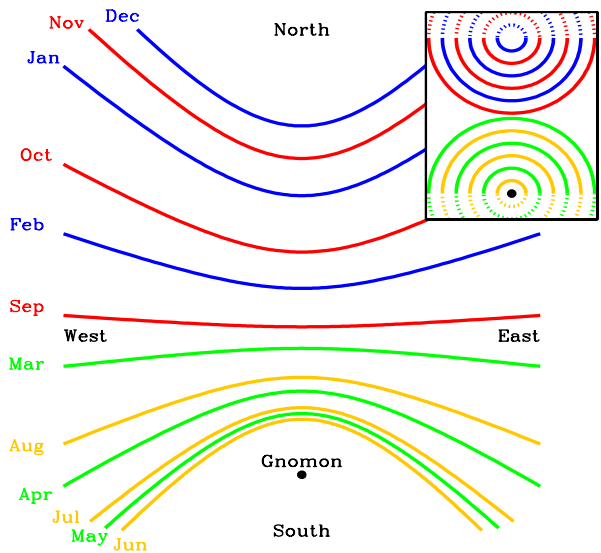

**Figure 4.** The curves of this gnomon pattern represent the daily path of the shadow of the tip of a vertical gnomon (black dot), cast by the sun on a horizontal ground (the figure plane). They are calculated for a typical latitude of 45° N on the 28th of each month using exact formulas (Demers & DuPuy, 1980). The colour coding is green, yellow, red, and blue for spring, summer, autumn, and winter, respectively. The upper-right figure represents a stylised idealisation of the observed pattern – including the necessary return of the sun from west to east over the night (dotted line) – as two sets of concentric circles.

While the two solstices are clearly distinct when living at high latitude – one is light and warm and the other is dark and cold – the two equinoxes are usually thought to be indistinct. Actually, there is a subtle difference that is easy to observe, but that does not affect people. The wide,

'dominant' arc of the summer sun is followed by the waxing, D-shaped half-moon (i.e. the first quarter moon) at the spring equinox, whereas at the autumn equinox it is traced by the waning, C-shaped half-moon (i.e. the third quarter moon). This intriguing difference is schematically illustrated in Figure 2. The ~5° inclination of the moon orbital plane will induce somewhat higher or lower paths of the moon compared to those of the sun, but the effect is small (at most at a 5°/23.5° = 21% level) and it averages out over decades. The result is that one can divide the year according to the shape of the dominant half-moon. The waxing half-moon is dominant from the winter to the summer solstice – the winter-spring semester – whereas the waning half-moon is dominant in the summer-autumn semester.

*The gnomon pattern*

A final interesting seasonal manifestation is the pattern of curves traced by the shadow cast by the tip of a gnomon. The stick at the centre of a sundial is a gnomon, which in its simplest form is a wooden post hammered in the ground. Figure 4 shows a series of curves traced by the shadow of a gnomon's tip every month over a year. This curve is bending around the gnomon at the summer solstice, flattens out towards the autumn equinox and bends again in the opposite direction during the autumn season around an imaginary point located further north. The pattern then inverses again from the winter to the summer solstice passing by the straight west-east line at the spring equinox. A plausible stylised representation of this yearly pattern could be in the form of two juxtaposed groups of concentric circles or a pair of spirals. This has striking similarities with the numerous cup-and-ring motifs found both in Ireland and Great Britain and their frequent association with spirals (van Hoek, 1993; Waddington, 1998). Figure 5 illustrates this by reproducing the motif of three stones (Brennan, 1983) found in archaeological sites far from the Boyne Valley and displaying diverse petroglyphs likely inspired by the gnomon pattern.

In summary, the main seasonal manifestations of the sun and the moon are:

1. the ratio of 12 or 13 lunar cycles (months) per solar cycle (year);
2. the increase of daylight in the winter-spring semester and its decrease in summer-autumn;
3. the associated widening, respectively shrinking, of the arc daily traced in the sky by the sun;
4. the large, 'dominant' path in the sky of the first quarter moon in the winter-spring semester and of the third quarter moon in summer-autumn;
5. the gnomon pattern with an equinoctial flip from a light spring-summer semester to a dark autumn-winter.

In addition, a calendrical representation would also aim to represent the endless succession of years by having the start and the end point of a linear motif at the same location. The engraved line can thus be followed without lifting the finger from one year to the next.

From a more pictorial point of view, the seasonal variation of the daily path of the sun in the sky draws two complementary motifs: a spiralling pattern (Figure 3), and the gnomon pattern (Figure 4).

**Calendrical interpretation of five spiral motifs**

After this long, but necessary presentation of seasonal celestial manifestations, we now focus our attention on five specific spiral motifs from the simplest to the most elaborated one. We will show that they seem to incorporate an increasing number of these annual manifestations, thus suggesting that they are calendrical symbols.

We start with the great spiral on the chamber stone C3 inside the west recess of Newgrange (Figure 6). The spiral widens counter-clockwise, which is the direction of the monthly movement of the moon against background stars. It traces 13 loops, which could therefore represent 13 moon revolutions around the earth in a year. There are many arguments in support of a simple calendar made of 13 x 28 days, falling just one day short of a year of 365 days (see above). The four arcs around the spiral could possibly represent the four quarters of the moon phases exactly as our seven-day weeks, since 4 x 7 days completes a month of 28 days. The two short arcs could thus be stylistic representations of the moon crescents of the first and last quarters and the two long arcs of the gibbous waxing and waning moon of the second and third quarters.

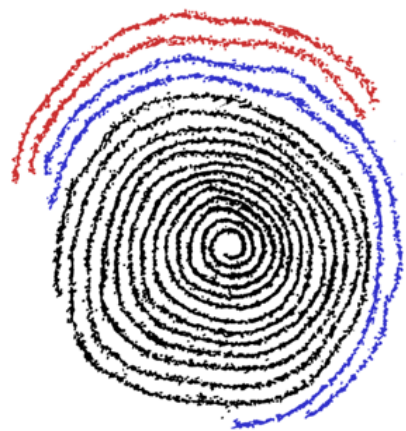

**Figure 6.** The great spiral on stone C3 in the west recess of Newgrange (from O'Kelly, 1982). The engraved motif is composed of a spiral with 13 turns widening counter-clockwise surrounded by two long arcs (shown in blue) and two short arcs (in red). The motif could represent a calendar with 13 months of exactly four weeks each.

Two other prominent spirals display a count of 13 cycles (Figure 7). They are more complex and seem to include three additional calendrical features: 1) the endless succession of years, 2) the increasing daylight duration from the winter to the summer solstice, and 3) the associated widening of the sun's daily path in the sky.

These features are apparently all incorporated in the spiral engraved, interestingly, just next to what is likely a sundial on kerbstone K15 in Knowth. The motif starts from the centre like the 13-loop spiral, but subtlety returns on the same path after seven widening turns to spiral back into the centre over six additional rotations. Assuming each turn to represent a month with the winter solstice at the centre, the seven widening and six shrinking turns could represent the variations over months of the arc traced by the sun in the sky (see Figures 2 & 3). The associated longer daylight in the summer, would then also be represented by the longer duration needed to engrave – or follow with the finger – the outer loops. With the year both starting and ending at the central point, it is possible to follow the groove over many years without lifting the finger.

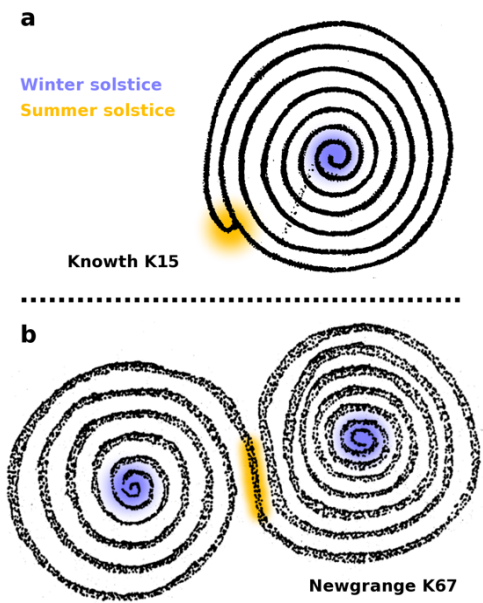

**Figure 7.** Two other spirals with a count of 13 loops. **a.** The motif on kerbstone K15 at Knowth (from Brennan, 1983) spirals out for seven turns before turning around to spiral in on the same line for six additional rotations. **b.** The S-shaped spiral on kerbstone K67 at Newgrange (O'Kelly, 1982) can be seen as an unfolded version of the same pattern. Both motifs could represent a year of 13 months with the winter solstice at the centres of the spirals (blue spots) and the summer solstice at the rewinding points (yellow shades).

All three calendrical features are equally present in an S-shaped spiral in Newgrange on kerbstone K67, which is apparently an unfolded version of the previous motif. The 13 loops are now disposed in order to divide the year into winter-spring and summer-autumn semesters with a transition at the summer solstice. This fundamental annual event could have been marked by placing in front of the kerbstone a wooden post or a now-lost standing stone, which would have shed its shadow in between the two spirals at the summer solstice. This would work according to the north-eastern position of this richly decorated stone (Brennan 1983:192) and would be analogue to the standing stone in front of the west passage of Knowth (see above).

A more elaborated spiral pattern is located in the chamber of Newgrange on stone C2 just to the left of the simple 13-loop spiral of Figure 6. As shown in Figure 8, the engraving of two intertwined spirals leaves in relief a continuous path that is like a rope wound from its middle, producing a 'double-spiral' with each extremity tracing a distinct path towards the centre. The symbolic importance of this motif is reflected by the presence of five such spirals on the entrance kerbstone K1. The motif is similar to the returning spiral on the stone K15 in Knowth (Figure 7), except that the inward and outward spiralling paths are now distinct. The previous calendrical features are present with the winter solstice being again supposed to be at the centre of the spiral and the summer solstice at its border. The motif enables an endless succession of years thanks to the entry and exit points next to each other on the left-hand side. However, this characteristic implies a non-integer number of loops. Assuming a calendrical representation, keeping an integer number of months in a year is only possible with a slightly modified way of counting them. An interesting possibility when following the spiral path with the finger is to count a month each time the orientation of the traced arc matches the crescent of the 'dominant' half-moon, which we have seen above to be the waxing, D-shaped moon in the winter-spring semester and the waning, C-shaped moon in the summer-autumn. Doing this with the slightly inclined orientation – at such a northern latitude – of this 'dominant' moon shape as seen rising on the east horizon results in a count of 12 months for the full pattern. The five filled lozenges of the chessboard pattern at the top of the spiral could tentatively represent the missing days to complete a year of 12 x 30 days, while with the addition of the six empty lozenges, they could represent the 11 days missing to complete the year after 12 synodic months of 29.5 days.

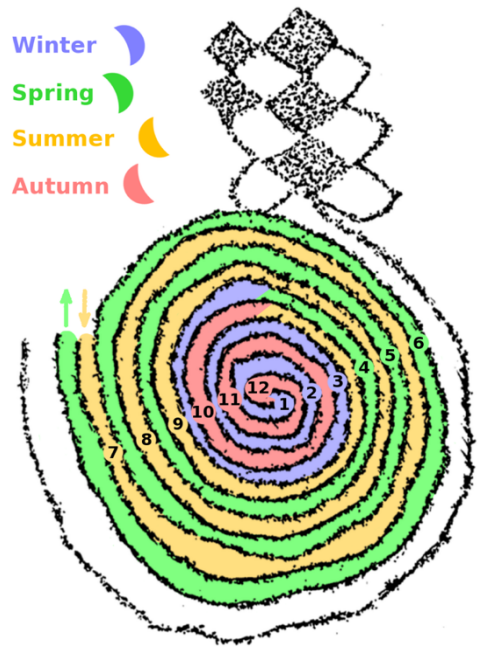

**Figure 8.** The double spiral motif on the chamber stone C2 inside Newgrange (from O'Kelly, 1982). The engraved motif leaves in relief a wound 'double-spiral', which is coloured in blue, green, yellow and red to mark the supposed winter, spring, summer and autumn seasons, respectively. The corresponding numbering of the months from the summer solstice (yellow arrow) to the winter solstice in the centre and back to the summer solstice (green arrow) is indicated with numbers (January=1). A month is supposed to be counted each time the spiral arc matches the orientation of the crescent shape of the 'dominant' half-moon as seen rising on the east horizon and represented schematically in the upper-left corner of the image.

An even more elaborated spiral pattern is located on kerbstone K13 at Knowth (Figure 9). The central autumn-winter part of the previous 'double-spiral' is kept, but the outer spring-summer part is wrapped around another centre. A total of 12 months is obtained when counting them, as before, each time the bend of the spiral corresponds to the orientation of the 'dominant' half-moon as seen when it is rising on the east horizon. The intention of this additional feature seems to be to separate the year into a light spring-summer semester and a dark autumn-winter with a transition at the equinoxes. The two distinct centres at the solstices and the change of bend through a straight line at the equinoxes is perfectly matching the observations of the shadow of a gnomon over the year (see Figure 4). It suggests that this observation has led to a rethinking of the calendrical representation to incorporate the important transition occurring at the equinoxes. That this motif is engraved at Knowth is in accordance with the roughly equinoctial alignment of the passageways of this mound.

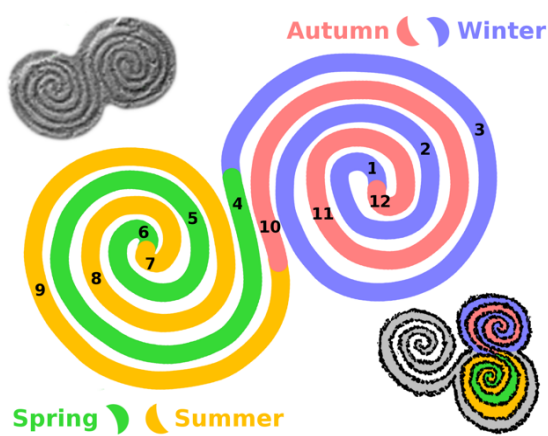

**Figure 9.** The delicate S-shaped double spiral on kerbstone K13 at Knowth. The central motif uses the same colours as in Figure 8 to indicate the proposed seasons and is based on the exact shape of the engraved pattern as derived via 3-dimensional capture and shown in the upper-left corner (Corns et al., 2015, www.3dicons.ie). Here again, the suggested count of months corresponds to the orientation of the rising half-moon visible on the east horizon and starts with the winter solstice (January=1). The lower-right image shows that the same motif forms a part of the beautiful triple-spiral in the central recess of Newgrange (from O'Kelly, 1982).

**Discussion**

*On the validity of the calendrical interpretation*

The presented interpretation is speculative by nature and its validity cannot be proven. The assumptions are: 1) that the signs are not purely decorative, but have some meaning; 2) that some of them are most likely related to seasonal celestial manifestations; and 3) that their meaning can be read from their form. Several arguments supporting the two first assumptions are presented in the introduction, whereas the third is inherent to the chosen approach. We live under the same sky as the Neolithic people some 5'000 years ago and have the same cognitive abilities. They could easily have observed the seasonal manifestations of the sun and the moon listed above. Our interpretation suggests an intentional link between these observations and specific spiral designs engraved on the rock. There is no certainty that this is correct, but there are many clues on five different stones that make this plausible.

A way to test the interpretation is to ask ourselves, whether there are more convincing calendrical representations than those presented here. Is it possible to design better motifs to represent month cycles in a year? Nowadays, we tend to represent a cyclical process with a sinusoidal line, but the succession of a series of circular cycles is more naturally rendered with a spiral motif. A spiral of 12 or 13 turns is therefore a very plausible representation of 12 or 13 moon cycles in a year. More explicit calendrical representations would be the spiralling yearly path of the sun on the sky (Figure 3) or the gnomon pattern (Figure 4). These motifs are currently not really part of our culture, but would actually be the most universal representations of a year. There is no known rock art precisely reproducing them, but we have shown above that some petroglyphs have striking similarities with them. Rather than accurately reproducing the observed yearly patterns, the artists apparently took the liberty to modify them into simple, elegant designs. For instance, the crossing of the enlarging spiral by the diminishing one in Figure 3 was apparently deliberately avoided when designing the double-spiral motif of Figure 8, although this simplification causes an alternation of the direction of rotation every semester. This is very interesting, because it tells us that the aim was apparently not an accurate, 'scientific' description of the celestial manifestations. The sky would rather have been used as a source of artistic and cultural inspiration.

*The five spirals put in context*

We considered all elaborated spiral motifs to be found either inside or outside the mounds of Newgrange and Knowth and present five spirals among them. There are many more engraved spirals in Knowth, but usually they have a much simpler design and therefore are not suited for the kind of detailed individual analysis presented here. The same applies to Dowth, the third great mound of the Bend of the Boyne, which has no prominent spiral motif. A statistical study is beyond the scope of this article and is not easy to conceive given the type of interpretive analysis of this paper. One could count the number of turns in either clockwise or anti-clockwise coiled spirals, but the conclusions that can be drawn from this are likely rather limited.

There are, however, several other complex spiral motifs that are not discussed. Most notably the famous triple-lobe spiral delicately engraved in the central recess of the chamber of Newgrange on stone C10 (e.g. O'Kelly, 1982; Stout & Stout, 2008: 40). Very interestingly, however, two of the lobes have precisely the same pattern as the S-shaped double-spiral of Knowth (Figure 9). The only difference is that the motif is picked in relief like the 'double-spiral' on stone C2 (Figure 8) and is connected to a third spiral on the left. The implications of this finding and the discussion of this beautiful sign, which could have had a symbolic power (Stout & Stout, 2008), shall be addressed in a separate article. Other omissions are the five 'double-spirals' magnificently engraved on the entrance stone K1 of Newgrange. We note, however, that the group of three spirals on its left side appears to be a stylistic version of the three-lobe spiral on stone C10, possibly with the intention to refer to it without exposing its exact pattern to the view of everybody. The two other double-spirals on kerbstone K1 and the pair on kerbstone K52 located on the other side of the mound could then also just be suggestive representations of the more genuine patterns discussed here. Their appearance in adjacent pairs is noteworthily reminiscent of the gnomon pattern like the examples in Figure 5 and many others, in particular at Loughcrew (Brennan, 1983: 180-182). Furthermore, the spirals on the 'calendar stone' K52 at Knowth and on the adjacent stone K51 are actually also 'double-spirals' and according to the count of months proposed in Figures 8 and 9 would represent a six-month semester. Therefore, although we presented only five examples, other spirals show similar characteristics and are rather corroborating than disproving our interpretation.

*Relation to earlier work*

Earlier attempts to relate the engraved symbols with astronomical cycles did not identify spirals as the most likely shape to represent them. Brennan (1983: 188) correctly noted that the sun traces a clockwise spiral with daily turns increasing in size from the winter to the summer solstice on the southern horizon and concluded that clockwise spirals are representations of the sun. He also noted a link between the gnomon pattern (see Figure 4) and an S-shaped spiral, but this relation was deduced from the work of an American artist, Charles Ross, who used a magnifying glass to focus the sun rays every day over a year to burn curves on wooden planks subsequently assembled to form a double spiral (Brennan, 1983: 190). He did not push the argument further to recognise that S-shaped spirals could be calendrical representations.

Thomas (1988) identifies an S-shape spiral as a symbol of the equinox linking the winter sun (a clockwise spiral) to the summer sun (an anti-clockwise spiral), but without justifying this interpretation. He does not mention the link with the gnomon pattern when discussing the double S-shaped spiral of Knowth on kerbstone K13, but had apparently the right intuition when stating: "It is possible the left and right spirals may represent the summer and winter solstice, the centre pair of joined spirals the equinox events." (Thomas 1988: 41).

For completeness, one should also mention that the fan-shaped motif on kerbstone K15 at Knowth was interpreted as representing a calendar, instead of a sundial (Thomas, 1988; MacKie 2009; 2013). According to these authors, it represents the prehistoric sixteen-'month' solar calendar, which was first proposed by Thom (1967) based on a statistical study of many standing stone alignments. The idea is to subdivide the year into halves, quarters, eighths and finally sixteenths leading to 'months' of 21 to 23 days, which are thus unrelated to the moon cycle, whereas most calendars are lunar or lunisolar (Richards, 1998). Actually, Thom's and MacKie's arguments for such a prehistoric calendar were criticised by Ruggles & Barclay (2000).

**Conclusion**

We successfully interpreted five spiral petroglyphs as possible calendrical representations. Although the interpretation is speculative and its validity cannot be proven, there is a beam of clues giving us confidence that we may have found a key towards understanding a small part of the Irish rock art. This key is a new approach, which we call 'dynamic', because it no longer considers the petroglyphs as static pictures, but adds a temporal dimension to them. The time aspect of the motifs can be apprehended by considering the engraving process by the artist or simply by following the patterns with the finger.

The coherent picture that emerges suggests that at least one or a few individuals of the Neolithic community of the Boyne Valley developed skills to observe astronomical cycles and to represent them by spiral motifs. The increasing complexity of the designs seems to follow the wish to include more observables and thus to have a more complete description of the celestial manifestations. This full process – from careful observations to imagining specific designs and engraving them for preservation – looks like a scientific methodology. The liberty taken by the artists to not simply reproduce the celestial manifestations suggests, however, that the original motivation is not scientific. The artist's intention could have been to encapsulate the knowledge of the cosmos in powerful symbols for a ritual purpose.

**Acknowledgements**. The author thanks Sonja Draxler for her work as organiser of the SEAC 2018 conference in Graz and editor of these proceedings, as well as the anonymous referees for helpful comments. He is grateful to Martin Brennan for his pioneering study of 1983 and his compilation of beautiful drawings, which were a prime source of inspiration after visiting Newgrange in 2010.